\documentclass{aa}
\usepackage{graphics}

\newcommand{\lsim}{\mathrel{\rlap{\lower4pt\hbox{\hskip1pt$\sim$}}\raise1pt\hbox{$<$}}}

\begin{document}      

   \title{The precision of large radio continuum source catalogues}
   \subtitle{An application of the SPECFIND tool}

   \author{B.~Vollmer\inst{1,2}, E.~Davoust\inst{3}, P.~Dubois\inst{1}, F.~Genova\inst{1}, F.~Ochsenbein\inst{1}, W.~van~Driel\inst{4}}

   \offprints{B.~Vollmer, e-mail: bvollmer@astro.u-strasbg.fr}

   \institute{CDS, Observatoire astronomique de Strasbourg, UMR 7550, 11 rue de l'universit\'e, 
     67000 Strasbourg, France \and
     Max-Planck-Institut f\"ur Radioastronomie, Auf dem H\"ugel 69, 53121 Bonn, Germany \and
     UMR 5572, Observatoire Midi-Pyr\'{e}n\'{e}es, 14 avenue E. Belin, 31400 Toulouse, France \and
     Observatoire de Paris, Section de Meudon, GEPI, CNRS UMR 8111 and Universit\'e Paris 7, 
     5 place Jules Janssen, 92195 Meudon Cedex, France
   } 
          
   \date{Received / Accepted}

   \authorrunning{Vollmer et al.}
   \titlerunning{Precision of large radio catalogues}

\abstract{
The accuracy in position and flux density of 19 large radio continuum
source catalogues has been determined using SPECFIND, a new tool
recently made available through the CDS. The $\sim 67\,000$ {\rm radio continuum} 
spectra with three or more frequencies produced by
SPECFIND were used to cross-correlate sources from different catalogues
and to calculate offsets in right ascension and declination
in the various catalogues with respect to the positions given
in the NRAO VLA Sky Survey (NVSS) catalogue, {\rm which was} adopted as a reference.  
The flux densities reported in the
catalogues were compared to those predicted by the {\rm composite} spectra,
enabling us to assess the quality of the flux density calibration
of the different catalogues.
}

\maketitle

\section{Introduction \label{sec:intro}}

A basic, fundamental source of information for the evaluation and use of large radio continuum 
surveys is the position and flux density accuracy of the final source catalogues. 
The precisions in flux densities and positions can be determined through the
examination of several sources of {\rm potential} errors.
We distinguish between direct errors connected to observations and errors due to the extraction of sources 
from observed maps. The main sources of uncertainties are:
\begin{itemize}
\item
Position: telescope pointing errors, which can amount to $\lsim 10$\,\% of the HPBW,
as well as errors introduced by the algorithm for source extraction from the observed maps, 
i.e. usually the fitting of a two-dimensional Gaussian. 
\item
Flux density: errors in the absolute flux calibration (see, e.g., Ott et al. 1994),
which are usually at the $\lsim 10$\,\% level, as well as errors introduced by
the source fitting algorithm -- especially for extended, non-Gaussian sources,
Gaussian fits can be insufficient or lead to the fitting of artificial multiple components;
it is very difficult to obtain error estimates of the latter effects.
\end{itemize}

To determine the position accuracy of large radio source catalogues, these are in general 
cross-correlated with other published radio source catalogues, which have a higher resolution
and which were generally obtained at different frequencies.
Often used reference catalogues are the  TeXas Survey of discrete radio sources (TXS, Douglas et al. 1996) 
and the Green Bank 4850~MHz survey (GB6, Gregory et al. 1996).
To determine the flux density accuracy of catalogues, 
there are two possibilities: if there is another published 
catalogue at the same frequency, the flux density scales can be compared directly,
and if there are (at least) two other suitable catalogues at different frequencies, the slope of the 
radio continuum spectra can be calculated based on these catalogues and the flux density
{\rm at the newly observed frequency} can then be inter- or extrapolated.

In a recent paper (Vollmer et al. 2004), we presented the new SPECFIND tool for the extraction
of cross-identifications and radio continuum 
spectra from radio source catalogues contained in the VizieR database of the Centre de Donn\'ees
astronomiques de Strasbourg (CDS).
The SPECFIND code can handle radio source catalogues made at different frequencies with different resolutions.
For the radio spectrum a power law is assumed, which can have one break if more than two
independent frequency points contribute to each of the two slopes.
Since our cross-identification is based on the comparison of flux densities at the same
frequency and on the radio spectrum at different frequencies, SPECFIND takes into account
whether a source is resolved by a given survey or not -- if, e.g., a source
is resolved only by our reference survey, the NRAO VLA Sky Survey (NVSS; Condon et al. 1998),
and not by other available surveys, there will be no
positive cross-identification, because the flux density of the resolved NVSS source
cannot be fitted into the radio spectrum of the unresolved source as defined by the other catalogues.

With this tool we identified $\sim 67\,000$ radio continuum spectra with more than two
independent frequencies {\rm between 178 and 8400 MHz} over the whole sky. 
With these spectra it is now possible to assess
the position and flux accuracy of the 19 large survey catalogues that we used for the
spectrum determination (Table~\ref{tab:entries}). Since the 1400~MHz NVSS 
is a complete, deep, interferometric survey, with by far the largest number ($\sim 1.8$ million)
of sources among these catalogues, we decided to use it as the reference for the positions 
of the other catalogues.

\begin{table}	
      \caption{SPECFIND radio continuum source catalogue entries.}
         \label{tab:entries}
      \[
         \begin{array}{lrrrrr}
{\rm name} & {\rm frequency} & {\rm resolution} & {\rm S}_{\rm min} & {\rm number\ of} &  {\rm Reference} \\
 & {\rm (MHz)} & {\rm (arcmin)} & {\rm (mJy)} & {\rm sources}  & \\
\hline
{\rm JVAS} &  8400 &  0.0055   & 30 & 2246 & (1)  \\
{\rm GB6}  &  4850 &  3.5      &  18 & 75162  & (2) \\
{\rm 87GB} &  4850 &  3.5      &  25 & 54579 & (3) \\
{\rm BWE}  &  4850 &  3.5      &  25 & 53522  & (4) \\
{\rm PMN}  &  4850 &  3.5      &  20 & 50814  & (5) \\
{\rm MITG} &  4850 &  2.8      &  40 & 24180 & (6) \\
{\rm PKS}  &  2700 &  8.0      &  50 & 8264  & (7) \\
{\rm F3R} &  2700 &  4.3      &  40 & 6495 & (8)  \\
{\rm FIRST} & 1400 &  0.08333  &  1 & 811117 & (9) \\
{\rm NVSS}  & 1400 &  0.75     &  2 & 1773484 & (10) \\
{\rm WB}   & 1400 & 10.       & 100 &  31524   & (11) \\
{\rm SUMSS} &  843  &  0.75     &  8 & 134870 & (12)  \\
{\rm B2}  &  408  &  8.0       & 250 &  9929 & (13)  \\
{\rm B3} & 408  &  5.0      &  100 & 13340 & (14) \\
{\rm MRC} & 408 &  3.0      &  700 & 12141 & (15)     \\
{\rm TXS}  &  365  &  0.1      &  250 & 66841 & (16) \\
{\rm WISH} & 325  &  0.9      &  10 & 90357 & (17) \\
{\rm WENSS} & 325  &  0.9      &  18 & 229420 & (18)\\
{\rm MIYUN} & 232  &  3.8      &  100 & 34426 & (19) \\
{\rm 4C}  &  178  &  11.5     &  2000 & 4844 & (20)  
\end{array}
      \]
\begin{list}{}{}
\item
(1) Patnaik et al. (1992); Browne et al. (1998);\\ Wilkinson et al. (1998)\\
(2) Gregory et al. (1996)\\
(3) Gregory \& Condon (1991)\\ 
(4) Becker et al. (1991)\\
(5) Wright et al. (1994; 1996); Griffith et al. (1994; 1995)\\
(6) Bennett et al. (1986); Langston et al. (1990);\\ Griffith et al. (1990; 1991)\\
(7) Otrupcek \& Wright (1991)\\
(8) F\"{u}rst et al. (1990)\\
(9) White et al. (1998)\\
(10) Condon et al. (1998)\\ 
(11) White \& Becker (1992) \\
(12) Mauch et al. (2003)\\
(13) Colla et al. (1970; 1972; 1973); Fanti et al. (1974)\\
(14) Ficarra et al. (1985)\\
(15) Large et al. (1991)\\
(16) Douglas et al. (1996)\\
(17) de Breuck et al. (2002)\\
(18) Rengelink et al. (1997)\\
(19) Zhang et al. (1997)\\
(20) Pilkington \& Scott (1965); Gower et al. (1967)
\end{list}
\end{table}

\section{Results \label{sec:results}}

To perform the cross-identifications (Vollmer et al. 2004) we used a least absolute deviation 
fit for the fit of the radio continuum spectra, 
which is more robust against outliers than a classical $\chi^{2}$ fit.
This method has the disadvantage, however, that the distribution of deviations is not Gaussian, but
has in general a strong central peak and more extended wings. To circumvent this problem we 
made $\chi^{2}$ fits instead of least absolute deviation fits to the frequency points
found by SPECFIND, which results
in a Gaussian distribution of deviations. 

\begin{figure}
\resizebox{\hsize}{!}{\includegraphics{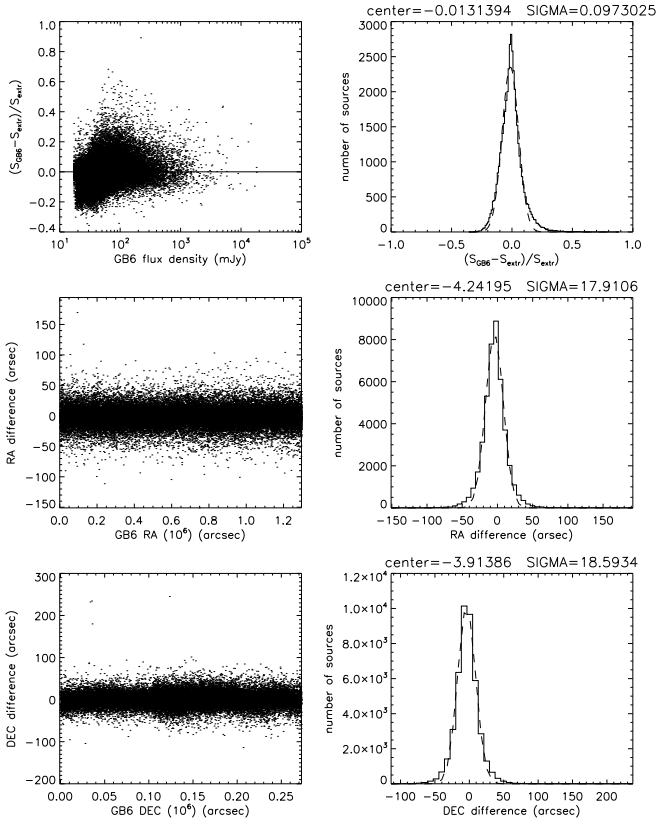}}
\caption{Positional and flux density accuracy of the GB6 survey with respect 
to the reference, which is the NVSS for position and the value predicted by
the composite spectrum from SPECFIND for the flux density.
Upper panel: flux density (see Eq.~\ref{eq:flux}), middle panel:
position offset in right ascension, lower panel: position offset
in declination. Right hand panels: solid line: observations, dashed line:
fitted Gaussian. The corresponding plots for the other catalogues can
be found in the online version of this article.} \label{fig:fig1}
\end{figure}

The calculation of the offsets
in the right ascension and declination of source positions is straightforward, and
the uncertainty in the flux density is measured as follows:
\begin{equation}
\Delta S = \frac{S_{\rm cat} - S_{\rm extr}}{S_{\rm extr}}\ ,
\label{eq:flux}
\end{equation}
where $S_{\rm cat}$ is the flux density in the catalogue and $S_{\rm extr}$ is the
flux density at the frequency of the catalogue as calculated using the fitted SPECFIND spectrum.
Fig.~\ref{fig:fig1} shows an example of the distribution of the differences in
flux density (upper panel), of the offsets in right ascension (middle panel) and
of the offsets in declination (lower panel). The corresponding plots for the other catalogues can
be found in the online version of this article. All three distributions can be fitted
by a Gaussian of form
\begin{equation}
N = N_{0} \exp\big(\frac{(\Delta - \Delta_{0})^{2}}{\sigma^{2}}\big)\ .
\label{eq:gauss}
\end{equation}

Since the accuracy of its position and the flux density of a radio source depend 
on the ratio of its flux density and the rms noise level of the survey, 
we calculate the uncertainties in the positions and flux accuracies for (i) all sources
in a given catalogue, sources whose flux densities are (ii) smaller than 3 times the minimum flux density of
sources in a given catalogue $S_{\rm min}$, (iii) between 3 and 10 times $S_{\rm min}$ 
and (iv) greater than 10 times $S_{\rm min}$. For sources near the catalogue {\rm sensitivity} 
the flux density uncertainty is typically $\sim 20$\%, whereas at high signal to noise ratios
the calibration uncertainty dominates.

For source positions and flux densities the derived values for $\Delta_{0}$ 
(the centre value, i.e., the mean systematic deviation
with respect to the reference, which is the NVSS for position and the value predicted by
the composite spectrum from SPECFIND for the flux density) and its standard deviation $\sigma$ 
are presented in Table~\ref{tab:spectab} for all catalogues that were included in
our analysis with SPECFIND (Vollmer et al. 2004), together with the accuracy in positions and
flux densities given by the catalogues' authors.
The columns are the following:
(1) catalogue name, (2) $\Delta_{0}$ for the right ascension, (3) $\sigma$ for the
right ascension, (4) $\Delta_{0}$ for the declination, (5) $\sigma$ for the declination
(6) positional accuracy according to the authors of the catalogue,
(7) $\Delta_{0}$ for the flux density, (8) $\sigma$ for the flux density and
(9) accuracy of the flux density according to the authors.

\begin{table*}	
      \caption{Positional and flux density uncertainties of selected radio continuum catalogues with
respect to the reference, which is the NVSS for position and the value predicted by
the composite spectrum from SPECFIND for the flux density.
}
         \label{tab:spectab}
      \[
         \begin{array}{|l|l|r|r|r|r||r||r|r||r|}
\hline
{\rm catalogue} & {\rm flux} & {\rm RA} & {\rm RA} &  {\rm DEC} & 
{\rm DEC} & {\rm position} & {\rm Flux} & {\rm Flux} & {\rm Flux} \\
 & {\rm range} & {\rm centre} & \sigma & {\rm centre} & \sigma & {\rm accuracy} & {\rm centre} & \sigma & {\rm accuracy} \\
 & (\times S_{\rm min}) & {\rm (arcsec)} & {\rm (arcsec)} & {\rm (arsec)} & {\rm (arcsec)} & {\rm (arcsec)} & \% & \% & \% \\
\hline
\hline
{\rm NVSS} & {\rm all} & - & - & - & - & <5 & -0.9 & 10.6 & 3 \\
 & <3 & - & - & - & - & & - & - & \\
 & 3-10 & - & - & - & - & & \sim 0 & \sim 10 & \\
 & >10 & - & - & - & - & & -0.9 & 10.5 & \\
\hline
{\rm JVAS} & {\rm all} & -0.1 & 0.5 & -0.2 & 0.5 & \ll 1 & -11.1 & 17.5 & 3 \\
 & <3 & - & - & - & - & & - & - & \\
 & 3-10 & -0.1 & 0.1 & -0.2 & 0.4 & & -12.6 & 15.3 & \\
 & >10 & -0.1 & 0.5 & -0.2 & 0.6 & & -7.2 & 21.8 & \\
\hline
{\rm GB6} & {\rm all} & -4.2 & 17.9 & -3.9 & 18.6 & 8 &-1.3 & 9.7 & 10 \\
 & <3 & -3.3 & 23.5 & -1.0 & 23.2 & & -2.1 & 9.3 & \\
 & 3-10 & -3.9 & 14.4 & -4.2 & 15.3 & & 0.3 & 10.7 & \\
 & >10 & -1.6 & 10.0 & -1.5 & 11.1 & & -0.2 & 10.2 & \\
\hline
{\rm 87GB} & {\rm all} & -3.2 & 19.9 & -3.9 & 22.9 & 10 & 8.5 & 14.5 & 14 \\
 & <3 & -3.5 & 24.6 & -3.7 & 26.6 & & 9.0 & 15.2 & \\
 & 3-10 & -1.7 & 16.5 & -3.1 & 19.7 & & 8.5 & 14.1 & \\
 & >10 & -1.4 & 12.5 & -2.8 & 16.7 & & 5.5 & 11.9 & \\
\hline
{\rm BWE} & {\rm all} & -1.6 & 19.5 & -2.7 & 22.4 & 20 & -4.7 & 11.7 & 10 \\
 & <3 & -1.7 & 23.7 & -2.7 & 25.6 & & -6.0 & 12.2 & \\
 & 3-10 & -0.8 & 15.1 & -2.0 & 18.4 & & -2.8 & 10.1 & \\
 & >10 & -0.8 & 12.7 & -1.9 & 15.8 & & -2.1 & 10.9 & \\
\hline 
{\rm PMN} & {\rm all} & -15.3 & 30.2 & -8.8 & 22.7 & <40 & -0.5 & 5.4 & 5 \\
 & <3 & -3.5 & 44.2 & -2.0 & 34.1 & & -0.5 & 4.7 & \\
 & 3-10 & -14.9 & 32.1 & -2.1 & 22.9 & & -0.3 & 5.1 & \\
 & >10 & -4.6 & 15.2 & -8.1 & 15.7 & & -1.2 & 8.9 & \\
\hline
{\rm MITG} & {\rm all} & -3.1 & 42.9 & -6.9 & 44.0 & 28 & -5.4 & 20.4 & 10 \\
 & <3 & -0.8 & 52.7 & -5.1 & 55.2 & & -8.6 & 19.4 & \\
 & 3-10 & -1.7 & 34.3 & -6.5 & 33.5 & & -2.0 & 18.1 & \\
 & >10 & 5.7 & 23.6 & -8.2 & 23.8 & & 3.0 & 19.1 & \\
\hline
{\rm PKS} & {\rm all} & -1.6 & 4.3 & -1.6 & 2.9 & 12 & -0.7 & 12.6 & >3 \\
 & <3 & -1.0 & 2.4 & -1.3 & 2.1 & & -6.1 & 15.4 & \\
 & 3-10 & -1.1 & 9.4 & -1.4 & 8.6 & & -1.1 & 12.6 & \\
 & >10 & -1.9 & 1.9 & -0.8 & 1.5 & & 0.7 & 11.4 & \\
\hline
{\rm F3R} & {\rm all} & 1.4 & 24.4 & -12.6 & 27.6 & 20 & 3.1 & 16.6 & 20 \\
 & <3 & 1.4 & 26.8 & -12.4 & 30.1 & & 2.2 & 17.9 & \\
 & 3-10 & 2.0 & 21.7 & -13.3 & 24.0 & & 4.0 & 15.0 & \\
 & >10 & 0.7 & 18.5 & -9.9 & 21.6 & & 4.5 & 12.1 & \\
\hline
{\rm FIRST} & {\rm all} & -0.5 & 0.6 & -0.2 & 0.3 & <1 & -2.5 & 13.1 & 5 \\
 & <3 & - & - & - & - & & - & - & \\
 & 3-10 & - & - & - & - & & - & - & \\
 & >10 & -0.5 & 0.6 & -0.2 & 0.3 & & -2.5 & 13.0 & \\
\hline
{\rm WB} & {\rm all} & -0.1 & 2.2 & -1.4 & 2.2 & 30 & 0.9 & 19.5 & 15 \\
 & <3 & -1.2 & 2.2 & -2.1 & 2.3 & & -0.9 & 20.7 & \\
 & 3-10 & -1.3 & 1.4 & -1.5 & 1.8 & & 2.7 & 17.2 & \\
 & >10 & -0.2 & 1.2 & -0.7 & 1.3 & & 4.8 & 12.3 & \\
\hline
{\rm SUMSS} & {\rm all} & -0.4 & 0.5 & -0.4 & 0.3 & 0.5 & 2.0 & 11.2 & 3 \\
 & <3 & - & - & - & - & & - & - & \\
 & 3-10 & - & - & - & - & & - & - & \\
 & >10 & -0.4 & 0.5 & -0.4 & 0.3 & & 2.0 & 11.1 & \\
\hline 
{\rm B2} & {\rm all} & -0.3 & 12.5 & -4.1 & 43.6 & 10/20 & 0.3 & 14.5 & 6 \\
 & <3 & -0.5 & 13.9 & -4.1 & 53.6 & & 0.6 & 15.4 & \\
 & 3-10 & 0.9 & 9.9 & -3.4 & 30.7 & & -0.4 & 12.5 & \\
 & >10 & -0.4 & 8.7 & -1.3 & 22.3 & & 2.7 & 11.8 & \\
\hline 
{\rm B3} & {\rm all} & -1.5 & 9.0 & -0.6 & 14.1 & <10 & -1.0 & 10.3 & 2 \\
 & <3 & -1.2 & 10.9 & -0.7 & 19.3 & & -0.2 & 11.0 & \\
 & 3-10 & -0.5 & 7.2 & -0.6 & 10.5 & & -2.3 & 9.3 & \\
 & >10 & -1.4 & 5.2 & -0.2 & 7.0 & & -1.6 & 7.9 & \\
\hline
\end{array}
      \]
\end{table*}

\addtocounter{table}{-1}
\begin{table*}	
      \caption{Uncertainties of catalogues with spectral indices (continued).}
         \label{tab:spectab1}
      \[
         \begin{array}{|l|l|r|r|r|r||r||r|r||r|}
\hline
{\rm catalogue} & {\rm flux} & {\rm RA} & {\rm RA} &  {\rm DEC} & 
{\rm DEC} & {\rm position} & {\rm Flux} & {\rm Flux} & {\rm Flux} \\
 & {\rm range} & {\rm centre} & \sigma & {\rm centre} & \sigma & {\rm accuracy} & {\rm centre} & \sigma & {\rm accuracy} \\
 & (\times S_{\rm min}) & {\rm (arcsec)} & {\rm (arcsec)} & {\rm (arsec)} & {\rm (arcsec)} & {\rm (arcsec)} & \% & \% & \% \\
\hline
\hline
{\rm MRC} & {\rm all} & -1.3 & 5.2 & -1.5 & 6.3 & 5 & 1.7 & 10.4 & 7 \\
 & <3 & -0.6 & 5.4 & -1.3 & 6.9 & & 1.7 & 10.5 & \\
 & 3-10 & 0.1 & 3.4 & -1.0 & 4.9 & & 1.6 & 9.6 & \\
 & >10 & -0.9 & 3.3 & -0.3 & 3.8 & & 3.9 & 11.7 & \\
\hline
{\rm TXS} & {\rm all} & -2.1 & 2.6 & -1.5 & 2.3 & 2 & 2.1 & 12.8 & 5 \\
 & <3 & -0.7 & 2.0 & -1.5 & 2.3 & & 2.2 & 13.4 & \\
 & 3-10 & -1.0 & 1.7 & -1.3 & 1.7 & & 1.5 & 11.8 & \\
 & >10 & -1.8 & 2.2 & -0.7 & 1.4 & & 3.3 & 10.9 & \\
\hline
{\rm WISH} & {\rm all} & -2.1 & 2.0 & -2.9 & 3.9 & >2 & -4.7 & 10.1 & 10 \\
 & <3 & - & - & - & - & & - & - & \\
 & 3-10 & -0.2 & 0.2 & -0.5 & 0.9 & & -0.2 & 9.8 & \\
 & >10 & -2.1 & 2.0 & -2.8 & 4.0 & & -5.0 & 10.0 & \\
\hline 
{\rm WENSS} & {\rm all} & -1.7 & 2.5 & -1.8 & 2.4 & >2 & 0.7 & 12.7 & 6 \\
 & <3 & -0.1 & 0.3 & -0.4 & 0.3 & & -0.9 & 10.9 & \\
 & 3-10 & -1.6 & 2.2 & -1.4 & 2.1 & & 1.0 & 13.0 & \\
 & >10 & -1.8 & 2.6 & -1.6 & 2.5 & & 0.7 & 12.7 & \\
\hline
{\rm MIYUN} & {\rm all} & -1.7 & 20.1 & -3.6 & 25.9 & >5 & -8.2 & 17.2 & 5 \\
 & <3 & -1.4 & 30.4 & -3.6 & 40.9 & & -7.5 & 16.7 & \\
 & 3-10 & -1.2 & 22.1 & -3.6 & 30.2 & & -9.8 & 17.1 & \\
 & >10 & -1.7 & 11.4 & -2.0 & 13.3 & & -6.1 & 16.6 & \\
\hline
{\rm 4C} & {\rm all} & -2.3 & 16.4 & -22.3 & 173.4 & 30/180 & -12.4 & 15.4 & 15 \\
 & <3 & -1.9 & 17.0 & -25.1 & 189.3 & & -12.3 & 16.1 & \\
 & 3-10 & -1.7 & 12.4 & -13.3 & 96.2 & & -12.4 & 12.9 & \\
 & >10 & - & - & - & - &  & - & - & \\
\hline
\end{array}
      \]
\end{table*}

\section{Discussion}

We assumed a single or double power law for the radio spectrum in the spectrum-finding
routine. Astrophysical sources show a single power law due to synchrotron emission in a 
frequency range which is limited by the domination of thermal emission by hot electrons at high 
frequencies (giving rise to a flat spectrum) and by synchrotron self absorption at low frequencies
(giving rise to a flattening or inversion of the spectral slope).
We take this into account by allowing a break in the radio spectrum
if the object appears in a sufficient number of catalogues of different frequencies.
For identification of two different slopes more than two independent frequency
points must be associated with each slope, which is not the case for the majority of
the sources. Therefore in general an over- or underestimate of the
flux densities is expected calculated using a single-slope SPECFIND spectrum
of sources with frequencies close to the edges of the frequency interval in which
a single-slope radio continuum spectrum is expected.

For the strong sources in the NVSS, we derived a positional error of about $1''$
from our comparison with the sources in the 8400 MHz JVAS catalogue
(Browne et al. 1998), which has by far the highest 
positional accuracy (milliarcsec) among those used in our study.
Note that the authors of the NVSS catalogue give positional uncertainties of $\lsim 1''$
for strong sources with flux densities greater than 15~mJy and of $7''$ for sources with
flux densities at the survey limit. All positional uncertainties given in Table~\ref{tab:spectab}
include this positional uncertainty of the NVSS.

The scatter of NVSS flux densities with respect to those expected based on the 
SPECFIND spectra is about 10\% for flux densities greater than 3\,$S_{\rm min}$. The centre of the
distribution of the JVAS flux density deviations is shifted to negative values,
due to the insensitivity of the JVAS observations (made in the VLA A configuration, which lacks short
baselines) to sources larger than $7''$, despite the possible flattening of the spectra towards 
8400~MHz. The relatively large scatter of the JVAS 
flux density deviations ($\sim 18$\,\%) is mainly caused by the VLA array parameters.

The systematic offsets in right ascension and declination are generally smaller
than $\sim 5$\,\% of the survey resolution for all catalogues with resolutions smaller
than that of the NVSS ($45''$). Moreover, all offsets are smaller than the catalogues' accuracy given by
their authors. 

Since the 87GB is based on a subsample used for the GB6, which is more sensitive,
we expect related results for their precisions in position and flux density.
Indeed, the positional accuracies are comparable, and the precision in flux
density is higher for the more sensitive catalogue (GB6), as expected.
The positional uncertainties given in the GB6 and 87GB catalogues
are for sources with flux densities greater than 10 times the minimum flux density
of the catalogue. For the B2 survey, the accuracy in declination is 
underestimated by $\sim 50$\,\% by the authors. For the MIYUN catalogue we 
also find a lower positional accuracy ($\sim 20''$ in right ascension and
declination for all sources and $\sim 10''$ 
for sources with flux densities greater than 10 times the catalogue flux limit) 
than the $>5''$ given by the authors.

There are no large systematic offsets in the flux densities per catalogue.
The flux density accuracy does not depend 
on the flux density of the sources for most of the catalogues (NVSS, JVAS, GB6, 87GB, BWE, PMN, MITG, PKS,
FIRST, MRC, TXS, WISH, WENSS, MIYUN).
The negative offsets of the 4C and MIYUN catalogues are most probably due to the flattening
or turnover of the spectrum around 200~MHz. In general, the flux density accuracy
lies in the range between 10\,\% and 20\,\% for all sources, and around 10\,\%
for sources with flux densities greater than 10 times the minimum flux density of the
catalogue. This seems to indicate that the effective calibration errors are
of the order of 10\,\% in all catalogues. The highest $\sigma$ value 
for all sources are found for the
JVAS, MITG, WB, and MIYUN catalogues. Our derived flux
density accuracies ($\sigma$) are consistent with those given by the catalogues' authors 
for about half of the catalogues (GB6, 87GB,
BWE, PMN, F3R, WB, MRC, WISH, 4C), whereas for the other catalogues we derived flux density
accuracies that are more than a factor of two larger than those given by the
authors. We did not detect a major problem in the flux density
scale of any of the catalogues included in this study.

This study shows that SPECFIND results can be used efficiently to assess the position
and flux accuracy of a large radio continuum catalogue. We therefore encourage
those responsible for future, as yet unpublished radio continuum catalogues to make
use of our code to optimise the position and flux density calibration of their surveys.
The SPECFIND master catalogue, the manual and the SPECFIND source files are publicly
available in the VizieR database at the CDS\footnote{Type SPECFIND in the 
VizieR front page: http://vizier.u-strasbg.fr/viz-bin/VizieR.}.

\begin{acknowledgements}
We would like to thank the anonymous referee for the suggestion to
divide the sources into several flux density bins. 
\end{acknowledgements}


\begin{thebibliography}{}


\bibitem{a3} Becker R.H., White R.L., \& Edwards A.L. 1991, ApJS, 75, 1 (BWE)

\bibitem{a4} Bennett C.L., Lawrence C.R., Burke B.F., Hewitt J.N., \& Mahoney J. 1986, ApJS, 61, 1 (MITG)

\bibitem{a5} Browne I.W.A., Patnaik A.R., Wilkinson P.N., \& Wrobel J.M. 1998, MNRAS, 293, 257 (JVAS)

\bibitem{a6} Colla G., Fanti C., Fanti R. et al. 1970, A\&AS, 1, 281 (B2)

\bibitem{a7} Colla G., Fanti C., Fanti R. et al. 1972, A\&AS, 7, 1 (B2)

\bibitem{a8} Colla G., Fanti C., Fanti R. et al. 1973, A\&AS, 11, 291 (B2)

\bibitem{a9} Condon J.J., Cotton W.D., Greisen E.W. et al. 1998, AJ, 115, 1693 (NVSS)

\bibitem{a13} de Breuck C., Tang Y., de Bruyn A.G., Rottgering H., \& van Breugel W. 2002, A\&A, 394, 59 (WISH)

\bibitem{a15} Douglas J.N., Bash F.N., Bozyan F.A., Torrence G.W., \& Wolfe C. 1996, AJ, 111, 1945 (TXS)

\bibitem{a17} Fanti C., Fanti R., Ficarra A., \& Padrielli L. 1974, A\&AS, 18, 147 (B2)

\bibitem{a18} Ficarra A., Grueff G., \& Tomassetti G. 1985, A\&AS, 59, 255 (B3)

\bibitem{a20} F\"{u}rst E., Reich W., Reich P., \& Reif K. 1990, A\&AS, 85, 805 (F3R)

\bibitem{a21} Gower J.F.R., Scott P.F., \& Wills D. MRAS, 1967, 71, 49 (4C) 

\bibitem{a22} Gregory P.C. \& Condon J.J. 1991, ApJS, 75, 1011 (87GB)

\bibitem{a23} Gregory P.C., Scott W.K., Douglas K., \& Condon J.J. 1996, ApJS, 103, 427 (GB6)

\bibitem{a24} Griffith M., Langston G., Heflin M. et al. 1990, ApJS, 74, 129 (MITG)

\bibitem{a25} Griffith M., Langston G., Heflin M., Conner S., \& Burke B. 1991, ApJS, 75, 801 (MITG)

\bibitem{a26} Griffith M.R., Wright A.E., Burke B.F., \& Ekers R.D. 1994, ApJS, 90, 179 (PMNT)

\bibitem{a27} Griffith M.R., Wright A.E., Burke B.F., \& Ekers R.D. 1995, ApJS, 97, 347 (PMNE)

\bibitem{a30} Langston G.I., Heflin M.B., Conner S.R. et al. 1990, ApJS, 72, 621 (MITG)

\bibitem{a30a} Large M.I., Cram L.E., \& Burgess A.M. 1991, The Observatory, 111, 72 (MRC)

\bibitem{a33} Mauch T., Murphy T., Buttery H.J. et al. 2003, MNRAS, 342, 1117 (SUMSS)

\bibitem{a45} Otrupcek R. \& Wright A.E. 1991, PASAu, 9, 170 (PKS)

\bibitem{a45a} Ott M., Witzel A., Quirrenbach A., et al. 1994, A\&A, 284, 331

\bibitem{a34} Patnaik A.R., Browne I.W.A., Wilkinson P.N., \& Wrobel J.M. 1992, MNRAS, 254, 655 (JVAS)

\bibitem{a35} Pilkington J.D.H. \& Scott P.F. 1965, MRAS, 69, 183 (4C)

\bibitem{a37} Rengelink, R. B., Tang, Y., de Bruyn, A. G., et al. 1997, A\&AS, 124, 259

\bibitem{a40a} Vollmer B., Davoust E., Dubois P., et al. 2004, A\&A, 431, 1177

\bibitem{a45d} White R.L., Becker R.H. 1992, ApJS, 79, 331 (WB)

\bibitem{a43} White R.L., Becker R.H., Helfand D.J., \& Gregg M.D. 1998, ApJ, 475, 479 (FIRST)

\bibitem{a44} Wilkinson P.N., Browne I.W.A., Patnaik A.R., Wrobel J.M., \& Sorathia B. 1998, MNRAS, 300, 790 (JVAS)

\bibitem{a46} Wright A.E., Griffith M.R., Burke B.F., \& Ekers R.D. 1994, ApJS, 91, 111 (PMNS)

\bibitem{a47} Wright A.E., Griffith M.R., Hunt A.J. et al. 1996, ApJS, 103, 145 (PMNZ)

\bibitem{a48} Zhang X., Zheng Y., Chen H. et al. 1997, A\&AS, 121, 59 (MIYUN)

\end{thebibliography}
\end{document}